NASA's Asteroid Grand Challenge: Strategy, Results, and Lessons Learned

**Jennifer L Gustetic**
[a]NASA Headquarters, 300 E St SW, Washington DC, 20546, USA,
jennifer.l.gustetic@nasa.gov, 904-504-8815

**Victoria Friedensen**
[a]NASA Headquarters, 300 E St SW, Washington DC, 20546, USA,
victoria.p.friedensen@nasa.gov

**Jason L Kessler**
[b]Shining Rock, LLC [c]79 Marlowe Rd, Fairview, NC 28730, USA,
jlkessler@me.com

**Shanessa Jackson**
[c]Valador, Inc [a]NASA Headquarters, 300 E St SW, Washington DC, 20546, USA
shanessa@nasa.gov

**James Parr**
[d]Director, Frontier Development Lab, 189 N Bernardo Ave #200, Mountain View, CA
94043, USA
james@frontierdevelopmentlab.org

Abstract

Beginning in 2012, NASA utilized a strategic process to identify broad societal questions, or grand challenges, that are well suited to the aerospace sector and align with national priorities. This effort generated NASA's first grand challenge, the Asteroid Grand Challenge (AGC), a large-scale effort using multi-disciplinary collaborations and innovative engagement mechanisms focused on finding and addressing asteroid threats to human populations. In April 2010, President Barack Obama announced a mission to send humans to an asteroid by 2025. This resulted in the agency's Asteroid Redirect Mission (ARM) to leverage and maximize existing robotic and human efforts to capture and reroute an asteroid, with the goal of eventual human exploration. The AGC, initiated in 2013, complemented ARM by expanding public participation, partnerships, and other approaches to find, understand, and overcome these potentially harmful asteroids. This paper describes a selection of AGC activities implemented from 2013 to 2017 and their results, excluding those conducted by NASA's Near-Earth Object Observations Program and other organizations. The strategic development of the initiative is outlined as well as initial successes, strengths, and weaknesses resulting from the first four years of AGC activities and approaches. Finally, we describe lesson learned and areas for continued work and study. The AGC lessons learned and strategies could inform the work of other agencies and organizations seeking to conduct a global scientific investigation with matrixed organizational support, multiple strategic partners, and numerous internal and external open innovation approaches and audiences.

# 1. Introduction[1]

In September 2009, the White House Office of Science and Technology Policy (OSTP) and the White House National Economic Council formally identified grand challenges as a key approach of the President's Strategy for American Innovation [1]. These challenges were defined as "ambitious but achievable goals that harness science, technology, and innovation to solve important national or global problems and that have the potential to capture the public's imagination [2]." The Obama Administration subsequently advised agencies across the US federal government of its commitment to increase the use of grand challenges as tools for promoting open government, innovation, and other national priorities: "We can set and meet grand challenges such as developing solar cells as cheap as paint, building anticancer drugs that spare healthy cells, and fitting the contents of the library of congress on a device the size of a sugar cube." [3].

NASA's first agency grand challenge was created in response to this strategy and unveiled in 2013. This paper summarizes the open innovation landscape that facilitated the early development of a model for grand challenges at NASA.  The selection, management, and implementation processes for the Asteroid Grand Challenge are outlined as well as key resources, stakeholders, and early results. The paper then describes several supporting projects that were initiated, followed by a detailed case study of a single noteworthy project. Finally, the authors share their assessment of the overall approach, while highlighting a series of areas for future study.

## 1.1 Background

The concept of grand challenges is not new–they have been used in some form by the private and non-profit sectors for over a century. David Hilbert famously outlined 23 mathematical problems at the Paris International Congress of Mathematicians in 1900, and challenged the world to solve them [3]. Later, this would inspire organizations in many fields to examine and identify major problems as grand challenges, such as the Grand Challenges in Global Health launched by the Gates Foundation in 2003 [4]. While some activities concluded at consensus building for the overall problems, such as the NIH-funded Grand Challenge for Public Health [5], some organizations took their investment in grand challenges further by making resources available for continued work. The Gates Foundation Grand Challenges in Global Health focused on 14 major scientific challenges by awarding 44 grants totaling over $450 million for research projects involving scientists in 33 countries [6]. Another famous example, the global effort to sequence the entire human genome, involved a $3.8B investment by the US. Federal Government over 13 years [7].

---

[1] Abbreviations
**AGC** – Asteroid Grand Challenge
**ARM** – Asteroid Redirect Mission
**CADET** – Citizen Science Asteroid Data, Education, and Tools
**DARPA** – Defense Advanced Research Projects Agency
**ECAST** – Expert and Citizen Assessment of Science and Technology
**HEOMD** – Human Exploration and Operations Mission Directorate (NASA)
**NEA** – Near-Earth Asteroid
**NEO** – Near-Earth Object
**NEOO** – Near-Earth Object Observations Program (NASA)
**NIH** – National Institutes of Health
**NPD** – NASA Policy Directive
**NTL** – NASA Tournament Lab
**OCT** – Office of the Chief Technologist (NASA)
**OSF** – Office of Strategy Formulation (NASA)
**OMB** – Office of Management and Budget (White House)
**OSTP** – Office of Science and Technology Policy (White House)
**PDCO** – Planetary Defense Coordination Office (NASA)
**PHA** – Potentially Hazardous Asteroid
**ROSES** – Research Opportunities in Earth and Space Science
**RFI**– Request for Information
**SAA** – Space Act Agreement
**SMD** – Science Mission Directorate (NASA)
**STMD** – Space Technology Mission Directorate (NASA)

These efforts were typically achieved through the large-scale investment of grants to researchers to accelerate focus and progress on these types of problems.

Earlier efforts, starting in the late 1990s, evolved the methods that could be applied to grand challenges. The Ansari XPRIZE and the DARPA Grand Challenge demonstrated that methods beyond grants, including incentive prizes, could also be used to stimulate progress against large problems. In 2009, the Obama Administration advised federal agencies of its commitment to increase the use of grand challenges as tools for promoting open government, innovation, and other national priorities. The Strategy for American Innovation stated, "We can set and meet grand challenges such as developing solar cells as cheap as paint, building anticancer drugs that spare healthy cells, and fitting the contents of the Library of Congress on a device the size of a sugar cube." [9]. The U.S. Department of Energy and the U.S. Agency for International Development (USAID) were the first federal agencies to start supporting grand challenge approaches in 2010 that used not only traditional grant making, but also incentive prize and crowdsourcing tools to engage "more than the usual suspects" in accomplishing the challenge goals.

However, the U.S. was not the only government during this time contemplating how grand challenges can influence and strengthen the foundation of a national science, technology, and innovation policy. A 2012 paper by Cagnin et al. explored how the European Union might reorient its innovation policy to "one that is more global in outlook and oriented towards so-called grand societal challenges" and described how portions of grand challenge approaches were being applied to the vision of Europe 2020 [8].

With a 60-year history of spinoff technologies (commercial products and services developed with agency assistance or investment), partnerships, new industries, and significant public interest resulting from various spaceflight and science programs, NASA is uniquely positioned to incorporate grand challenges as part of its strategic vision. This legacy has afforded various legal authorities, policies, and mechanisms, such as Space Act Agreements (SAAs) and prize competitions, which enable NASA to engage non-traditional partners and audiences with its mission. In response to these trends, NASA thus began a process in 2012 to explore how grand challenges related to space exploration could energize public and private constituencies around a clear "call to action" and spur considerable advancement in a wide range of domains. NASA anticipated that grand challenges could "serve as a 'North Star' for collaboration between the public and private sectors" [2] in areas where the Government alone is unlikely to achieve the outcome.

## 2. Selection Process for NASA's First Grand Challenge

NASA immediately responded to the Obama Administration's pursuit of grand challenges as a national innovation strategy. The Office of the Chief Technologist (OCT) initially coordinated the formulation of grand challenges as part of its responsibility to support the agency with principal advice and advocacy on technology policy and programs.

First, learning from other organizations' grand challenge approaches, OCT developed a conceptual model in 2012 of grand challenges as catalysts for unifying partners, unleashing creativity, and accelerating progress toward difficult and important science and technology goals. The details of this model became an important tool for describing to NASA how a grand challenge approach differs from the agency's standard way of doing business and spurs innovation in science and technology areas, for example by:

- Identifying a grand, unmet, and achievable human goal.
- Aligning that goal with national needs and priorities.
- Issuing a call to action to a diverse group of stakeholders.

- Facilitating coordination and collaboration among traditional and non-traditional entities on both existing and new actions.

Next, the NASA Executive Council (EC) determined in June 2012 that the agency should pursue the development of grand challenges. The council directed that agency-level grand challenges should be bold yet achievable goals in science, technology, and innovation, and that any proposed grand challenge must:

- Clearly relate to NASA's mission and strategy.

- Unify and inspire public and private constituencies around a clear "call to action". It must be specific enough to convey success criteria and broad enough to capture the imagination of participants and the public.

- Spur significant developments in a wide range of domains vital to the nation and the world.

- Serve as a beacon for high-impact, multi-disciplinary collaborations among the government, industry companies, universities, non-profits, philanthropists, and individuals including scientists and engineers. Given NASA's investments, partnerships and ability to leverage activities and resources, the grand challenge must be ambitious, but realistically achievable within 10-20 years.

Figure 1 illustrates many other factors NASA considered when selecting candidates for grand challenges, including suitability for a NASA leadership role, alignment with other federal agency goals, and the extent to which the grand challenge would leverage existing activities.

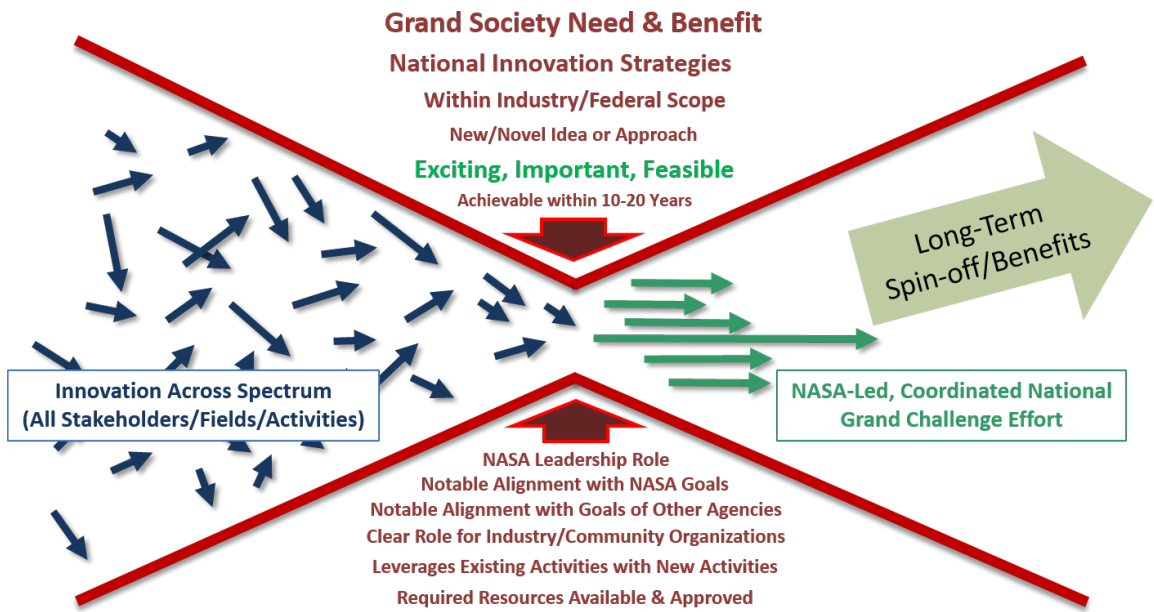

**Figure 1: Characteristics of NASA, Agency Grand Challenges**

Following this EC decision, NASA's OCT and Office of Strategy Formulation (OSF) led an effort from June 2012 through June 2013 to identify, formulate, refine, and select grand challenge topics appropriate for NASA to lead and coordinate. The OSF was responsible for agency-level policy efforts, facilitating the implementation of a wide range of initiatives in support of NASA's goals, and integrating the efforts of the agency's various strategic planning offices to ensure consistency with White House and legislative direction.

OCT and OSF concluded that selecting an inspirational and feasible grand challenge for NASA required the imagination and experience of a broad variety of experts. Preliminary concepts were gathered across NASA and synthesized in an inclusive process that culminated in a workshop at NASA Headquarters called a "Big Think". This workshop was designed as a generative session to test and examine possible challenge statements and determine rationale and logistics for each option ("the 'why' and 'how'") across several emerging themes, informed by the expertise of attendees. The Big Think revealed perspectives from senior NASA executives and external guests who were invited to provide a distinct point of view. Non-NASA participants included OSTP representatives, science fiction authors, and entrepreneurs.

Guided by the results of this workshop, NASA selected an Asteroid Grand Challenge (AGC) to "find all asteroid threats to human populations and know what to do about them" as its first grand challenge, through the Fiscal Year 2014 budget cycle in collaboration with the White House OSTP and Office of Management and Budget (OMB) [10]. Having met the characteristics described in Figure 1, the AGC was implemented with the intent to leverage existing activities, pursue new partnerships, and increase awareness of its compelling goal. The next section describes the structure of the AGC, activities initiated from 2013 to 2017, and some initial results from those activities.

## 3. Asteroid Grand Challenge Overview and Summary of Activities and Results, 2013-2017

In June 2013, NASA held a public Asteroid Initiative Industry & Partner Day, where the AGC was announced, and the forthcoming Asteroid Initiative was outlined [11]. Since 1998, NASA has led a global effort to coordinate the detection, tracking, and characterization of potentially hazardous asteroids (PHAs) and comets that could approach Earth. At the time, estimates suggested that NASA's Near-Earth Object Observations (NEOO) Program had surveyed less than 10% of objects smaller than 300 meters in diameter and less than 1% of objects smaller than 100 meters in diameter. It is understood that such objects represent a significant threat to Earth. A high-profile explosion and shockwave over Chelyabinsk, Russia in February 2013 dramatically illustrated the hazard represented by near-Earth objects (NEOs) as small as 30 meters in diameter [12]. This event reinforced the need for a global effort with innovative solutions to accelerate the completion of the survey of PHAs.

The AGC was a supporting element of the Asteroid Initiative [13], which was composed of two separate but related parts: ARM and the AGC. ARM was envisioned as a first-ever robotic mission to visit a large near-Earth asteroid, collect a multi-ton boulder from its surface, and redirect it into a stable orbit around the moon. The AGC complement ARM by using innovative partnerships and approaches to help us identify and track asteroids, while both efforts contend with the challenge of mitigation.

Observatories and organizations around the world coordinate extensively on finding and characterizing asteroid threats. With more than 95% of possibly devastating NEOs larger than one kilometer already discovered, the NEOO Program is focused on finding 95% of NEOs that are larger than 140 meters. Besides managing the detection and cataloging of NEOs, NASA's Planetary Defense Coordination Office (PDCO) has facilitated communications between the astronomical community, the federal Government, and the public in the event that potentially hazardous objects are discovered [14]. Other existing NASA efforts on this topic include partnerships with space institutes; other U.S. government agencies such as the Air Force, Federal Emergency Management Agency, and National Science Foundation; universities; and the space agencies of international governments [15]. Hazards from NEOs are understood as a preventable natural disaster; i.e., if detected early enough then something might be done to prevent an asteroid from impacting the Earth, or at the least, emergency planning could reduce the potential harm caused by an asteroid impact in the event there isn't enough warning. This concept, mitigation, is a key element of the PDCO's interagency coordination work and was considered a major topic for engagement with the public.

This paper does not describe existing work since the late 1990s of NASA's PDCO and NEOO Program, other U.S. Federal agencies, non-profit entities, or international organizations including the United Nations. Rather, the activities highlighted are those directly initiated as part of the AGC, in close coordination with the NEOO Program and PDCO.

During the first four years of the AGC, NASA inspired asteroid hunters and met citizen scientists, challenging people worldwide to get involved in protecting the planet from asteroid threats. From 2013-2016, the AGC was coordinated by OCT in close collaboration with the agency's Science Mission Directorate (SMD), Human Exploration and Operations Mission Directorate (HEOMD), and Space Technology Mission Directorate (STMD). In 2016, the responsibility for executing the AGC transferred to the PDCO in SMD.

It is important to note that NASA's AGC was intentionally designed to differ from other grand challenges, which focused on numerous grants to researchers (e.g. the human genome project) or a single incentive prize to solve the entire problem (e.g. DARPA's Grand Challenge). NASA did not intend to dedicate significant resources to the AGC, and traditional research grants were being implemented by SMD. Therefore, NASA's AGC efforts focused on augmenting those existing activities through *new* methods and activities applied to "find all asteroid threats to human populations and know what to do about them".

A wealth of project experience and literature have demonstrated how public engagement (including incentive prizes, crowdsourcing, citizen science, and public deliberations) provides meaningful contributions to science and technology problems. Additionally, NASA conducted numerous open innovation activities that successfully yielded promising results [16], which strengthened internal confidence in the AGC's expansion to new communities and approaches to address parts of the asteroid goa.

These new methods and activities fall within three categories:
- Awareness and Outreach
- Discovery and Follow-up Observations
- Characterization

The first category, awareness and outreach, is a new AGC construct that accounts for NASA's efforts to engage a broader "solver" community with filling gaps and accelerating work already underway. The second and third areas complement existing PDCO work to detect, track, characterize, and mitigate; with a focus on efforts deemed suitable for input by non-expert individuals who are not planetary scientists, which include follow-up observations and asteroid characterization.

## 4. Selected AGC-Initiated Project Summaries

This section provides highlights of a sampling of notable projects from the much larger body of AGC engagement and research efforts.

### 4.1 Awareness and Outreach

Based on the NASA EC direction for Grand Challenges, much thought was given to engaging audiences as widely as possible. Engagement approaches targeted traditional partners with common government tools as well as nontraditional groups such as the maker community, coders, and the general public, with the goal to respond actively to new ideas and approaches while implementing core activities.

As part of the June 18, 2013 Asteroid Initiative Industry Forum [13] during which the AGC was announced, NASA released a related Request for Information (RFI) [17] to seek new and innovative ways for individuals to engage with the agency. The AGC was created on the premise of a diverse solver community capable of accelerating NASA's existing NEO work, and the RFI release sought such ideas related to partnership and engagement activities. As a result, NASA received over 400 ideas for ARM and the AGC. In September and November 2013, NASA hosted public workshops in Houston to discuss 96 of the submitted ideas, including 11 that offered approaches for conducting outreach and engaging citizen scientists.

The RFI and the forum proved to be effective approaches for engaging traditional NASA constituencies; however, the responses were submitted from "the usual suspects" and generally did not include new voices. As NASA also sought to reach out more broadly, the agency became aware of a considerable lack of public understanding and readily consumable information related to asteroid threats. Therefore, the AGC refocused its first-year efforts on establishing an information resource about PHAs, engaging the public to understand their views of the challenge, and communicating the challenge directly to potential solver communities including NASA employees.

For example, to build the PHA information resource, NASA OCT and Ames Research Center's Solar System Exploration Research Institute (SSERVI) hosted a virtual seminar series in 2014 [18] to present the basics about near-Earth asteroids, while building an online resource for those interested in learning the fundamentals critical for contributing to the AGC. The eight seminars continue to be accessed from SSERVI's website, with more than 6200 total views as of January 2018 [19].

Also, after a presentation by the Expert and Citizen Assessment of Science and Technology (ECAST) network at the 2013 Asteroid Initiative workshop [20], NASA awarded a cooperative agreement to this consortium of universities, science centers, and non-governmental organizations. ECAST would conduct peer-to-peer deliberations and solicit citizen input on the Asteroid Initiative, including both the AGC and ARM. These innovative forums would provide the opportunity for NASA to meet its objective of engaging individual citizens for their input on the Asteroid Initiative.

Active in Europe, the method of utilizing participatory technology assessment (pTA) for engaging the public in facilitated policy conversations had never been tried by a U.S. government Agency. Given the mandate of the AGC to *unify and inspire public and private constituencies around a clear* "*call to action*", NASA saw this pTA activity as not only be appropriate for the AGC efforts but also as a potentially useful approach for other Agencies interested in engaging the public in new and meaningful ways.

In November 2014, ECAST hosted two in-person events in Boston and Phoenix and a virtual forum to gather input from the public, providing NASA with diverse, informed feedback about the Asteroid Initiative. Nearly 200 participants attended the in-person events, and 70 people registered for the virtual forum with 32 active participants.

The forums focused on three topics for deliberation: (1) planetary defense, (2) architecture options for ARM being considered by NASA, and (3) NASA's human exploration strategy. The results from the ARM architecture deliberation were presented to NASA leadership and were a part of the final determination and selection for the mission architecture [21]. Interim results on public deliberation related to planetary defense also inspired the PDCO to engage the European Space Agency (ESA) to develop follow-up forums to explore cultural and societal differences.

Of particular interest at the ECAST forums was strong support for a space-based telescope as a means of meeting congressionally mandated goals for finding threats to Earth, and nuanced acceptance of the use of nuclear explosive devices to disrupt an incoming large asteroid. Of note, this support for a space-based telescope was echoed by commercial partners at an AGC-convened two-day workshop entitled *The Economics of NEOs* in September 2014. At this workshop, over 100 participants from academia, the scientific community, international governments, aerospace agencies, commercial and industry organizations, and the public "isolated the most salient challenges and opportunities presented to the NEO

community… to better prepare [them] to meet the upcoming challenges and opportunities head on, [including the] need for a space-based telescope designed to be part of a broad survey detecting and characterizing asteroids for the benefit of all stakeholders in the NEO community [22]."

The final report covering all aspects of the ECAST engagement was shared with NASA and the public in August 2015 [23]. These forums resulted in many communications on public participation and the engineering process that were shared via conferences and publications, including the February 2017 American Association for the Advancement of Science Annual Meeting [24], the proceedings of the American Institute of Aeronautics and Astronautics SPACE 2015 [25], and the international journal Astropolitics [21]. The ECAST effort continues to inform the PDCO's efforts to develop government policies related to approaches for mitigating asteroid threats.

This model has been adopted broadly within U.S. government. For example, the ECAST team has supported the National Oceanic and Atmospheric Administration (NOAA) in a series of public deliberation forums to clarify public values around weather-related resiliency [26].

## 4.2 Discovery and Follow-Up Observations

With over 90% of NEOs larger than one kilometer already discovered, NASA is now focusing on finding 90% of the NEO population larger than 140 meters [27], while recent studies show that much smaller objects present statistically significant risks at local and regional levels. While large objects may be observable by amateurs, smaller NEOs are more difficult to find because progressively larger telescopes are needed to identify smaller and dimmer objects. However, this does not eliminate the role of non-professional observers in discovering new asteroids or contributing to follow-up observations of a newly discovered threat. Even without finding entirely new asteroids through direct observation, NASA recognized the vast volume of collected data could serve as a rich resource for discovery and follow up analysis by an amateur solver community. The AGC initiated a number of projects to address these issues, including a partnership with Planetary Resources, Inc. [28] to conduct the Asteroid Data Hunter Challenge [29], the Asteroid Data Tracker Challenge, student badging for asteroid hunting, and the Ultrascope Project. This summary will focus on the results of the badging and Ultrascope projects.

### 4.2.1 Student Badging

During the early development stages of the AGC, planners recognized student engagement as an important precursor for training future asteroid hunters. "Digital badging", analogous to badges earned by scouts, was considered a creative way to accomplish this objective, and a pilot program was created in 2015 to test the system, curriculum, and learning practices.  Six middle-school students underwent in an asteroid detection training program that badged them for skill acquisition through a commercial digital credential platform called Credly.  The students utilized data collected by professional asteroid survey teams to search for asteroids that may have been missed.  The pilot successfully concluded with 7th Graders from Dillard Drive School in Raleigh, North Carolina discovering 10 new asteroids.

Teaching materials were developed for the expansion of the program to educators worldwide.  During the summer of 2016, the first phase of the expanded program successfully engaged 90 schools in India as further demonstration of the concept [30]. This activity delivered outcomes for the awareness and outreach component of the AGC as well as the discovery category. While the initial foray was successful, personnel changes slowed the effort and NASA has not continued the project. However, this may be a promising approach for future engagement in discovery leveraging existing data.

### 4.2.2 Ultrascope

The Ultrascope project is a different example of how new methods and technologies could assist with discovery and follow-up observations. During the 2015 International NASA Space Apps Challenge, a team from the Open Space Agency (OSA) designed and built a robotically controlled, 3-D printed and laser-cut, "ultra low-cost" telescope called an Ultrascope. The project used a 6-inch mirror as a precursor model for a subsequent design for asteroid follow-up imaging and characterization using a 12-inch mirror. The vision of this project is to design an open source hardware kit-set robotic telescope and associated control software (running on smartphones) that anyone can build and refine, with the goal of providing a coordinated workflow for amateur astronomers to assist with follow-up observations.

Through the AGC, NASA signed an SAA in 2014 with SpaceGAMBIT, an organization funded by DARPA to spur aerospace innovation through seed funding for open-source projects. After the Space Apps Challenge, the OSA team was awarded one of ten grants from SpaceGAMBIT that enables them to improve upon the initial design of Ultrascope [31].

By using off-the-shelf parts and manufacturing techniques available to most schools and maker spaces, and leveraging the ever-increasing resolution and noise reduction capabilities of mobile phone cameras, the Ultrascope represents a 10x potential reduction in cost compared with commercial telescopes [32]. Additionally, the control software and cloud-based post-processing and control over a 4G network could further democratize the challenge of assisting asteroid characterization. These advancements could provide science educators and amateur astronomy networks across the globe a simplified pathway to establishing operational telescopes, including in locations where broadband is often unavailable, such as the Southern Hemisphere.

The aspiration of the Ultrascope project remains to lower the cost of precision astronomy and the often-challenging skill ramp-up associated with this kind of citizen science [33]. To resolve this second problem, the Ultrascope team also developed a 10-step badging process to assist new astronomers in both the build of their Ultrascopes and guide them to a successful asteroid light curve.

The Ultrascope's hardware and software files and plans have been released to the public as open-source projects. Ultrascopes are currently being built around the world - often with innovative modifications and improvements to original design. As of January 2018, more than 100 telescopes have been built globally and a community of more than 1,200 are following activities. Although sporting a 6-inch mirror, limiting its resolving power to main belt objects, the potential remains to provide an introduction to the need for asteroid characterization to students and citizen scientists and in turn, increase the number of individuals tackling the challenge of high-quality follow-up observations in the future. Plans are underway to design and develop the next generation 12-inch Ultrascope - which will be capable of viewing Near Earth Objects and contributing to the AGC in earnest.

One highlight of the project has been the Ultrascope assembled and deployed by students in a rural South African township which was remotely controlled by attendees in New York City at the 2015 Wired Business Conference - where live images of the Southern Sky where shown on a large screen. At a second venue, the RE:CODE conference in Los Angeles, another Ultrascope in Spain was used to stream an image of the Asteroid Vesta over 4G.

The Ultrascope remains a high impact engagement example of AGC activities. Over the duration of the project, more than 1.5 million individuals have been introduced to the Ultrascope and the need for asteroid follow-up observation. This has been amplified by films and demonstrations sponsored by Microsoft and Qualcomm, using the project to showcase the power of their new mobile phones and chipsets. The Ultrascope has been showcased at Maker Faires around the world, featured in numerous tech blogs and publications, demonstrated at a White House Astronomy Night [34], and featured in a dedicated Planetarium Show at the San Diego Maker Faire.

*4.3: Characterization*

Characterization of an NEO involves determining its physical characteristics, such as mass, density, porosity, and composition. During the creation of the AGC, managers assumed that the characterization category provided the greatest opportunity for external contributions through open innovation. Because expert-led surveys focus on detection and tracking, and request characterization for only a small subset of new discoveries, a gap remains in our understanding of NEAs that could be closed with mass contributions from the public. Light curves (the measurement of a celestial body's brightness at certain intervals and over a given period) and phase curves (the brightness of a reflecting body as a function of its phase angle) are both examples of important measurements typically collected by amateur astronomers. AGC-initiated projects that sought to increase capabilities for amateurs to contribute to characterization included the Citizen Science Asteroid Data, Education, and Tools (CADET) grant program [35], the Cube Quest Centennial Challenge [36, 37], and Apophis 2029. This summary focuses on the lessons from the Apophis 2029 project.

At the World Maker Faire in 2013, an unplanned discussion between the NASA AGC Program Executive and an advertising executive working with Verizon Wireless resulted in a two-year journey to create an easily accessible citizen science activity to provide a meaningful contribution to asteroid characterization.

Given the size and reach of the Verizon Wireless network, NASA sought to find a similar opportunity but with broader reach. Efforts focused on creating an engaging mobile game that also included a mechanism for people to help classify asteroids.

As early as 2011, new discoveries made on the citizen science platform Zooniverse through projects like Galaxy Zoo [38] and Planet Hunters [39] were being published in peer-reviewed journals. The applicability of the approach grew during subsequent years, and Zooniverse surpassed one million citizen science participants in 2014 [40]. To date, more than 55 articles in peer-reviewed science publications are related to activities conducted on Zooniverse [41]. This helped to establish that an understandable challenge on a well-designed platform can unlock the potential of thousands of interested participants and address scientific questions more quickly or beyond the original researcher's ability to complete alone.

After many months of brainstorming sessions, a presentation about the AGC to a handful of researchers at the Harvard-Smithsonian Center for Astrophysics led to a breakthrough. An MIT researcher, Francesca DeMao, used hyperspectral imagery to classify asteroids and maintained a database of over 50,000 uncharacterized asteroids (primarily main belt asteroids). With a problem identified that an extensive network could help solve, the next step was to develop an engaging mechanism to enroll people for their help. In 2015, NASA signed an SAA with Verizon to partner on this asteroid characterization work.

After many more months of game planning and design, leveraging the Verizon's budget t (which was larger than the entire AGC annual budget), an initial alpha version of the game, Apophis2029, was announced at the Cannes Lions Festival of Creativity in June 2015 [42]. Unfortunately, the partner organization experienced internal changes within its structure shortly after Cannes and headed in a new direction, so a beta version for external testing was not released.

This project was a good lesson in the power and risk of partnerships. The NASA AGC's limited budget would have been unable to support the creation of an alpha version of a game without the private sector partner. However, when the partner decided to move in a new direction, the game was not developed further, leaving an opportunity for asteroid characterization unrealized. As also illustrated by the multi-year development of the Ultrascope project, grand challenges open dialogue, offer possibilities, and create opportunities for all organizations and individuals to begin thinking about how they might support the effort.

## 5. Case Study: NASA Frontier Development Lab

The projects described in section 4 led to insights that informed a particularly effective cross-cutting AGC initiative: the Frontier Development Lab (FDL). Unveiled in 2016, FDL is an artificial intelligence (AI) applied-technology accelerator that engaged post-graduate researchers with tightly defined questions surrounding asteroid detection, characterization, and mitigation. Participants received significant support from industry technology partners and supportive feedback exchanges with planetary science and data science experts. FDL resulted from a conversation at a meeting of senior planetary scientists where a desire was expressed for the AGC to fund a fellowship program to help with current research gaps. The opportunity added an AGC nuance: to serve as a beacon for high-impact, multi-disciplinary collaborations among the government, industry, and academia. Based at the SETI Institute in Mountain View, California, FDL has received financial and in-kind support from leading companies in artificial intelligence, including Nvidia Corporation, Intel Corporation, IBM, and Google LLC. This project answers AGC's original goal of exploring non-conventional partnerships and platforms for private sector engagement.

FDL aims to provide researchers with meaningful research opportunities that directly complement the work of the planetary defense community, focusing on vigorous and accelerated pursuit of specific capability gaps. The first cohort in 2016 consisted of three interdisciplinary teams of four post-graduate professionals who joined mentors and guest lecturers in efforts to define an asteroid science problem and combine the skills of planetary science and machine learning to advance problem solving.

Results from the first cohort's activities include:
- A demonstration on how a drone equipped with machine vision could search for recently fallen meteorites (with a false positive rate of only 0.7 percent) and a tool for enabling citizen meteorite searchers to adapt drones similarly [43].
- A neural network approach (Variational Auto Encoder) and Bayesian Optimization to accelerate the process of 3-D asteroid radar shape modeling. Estimates of shape, center of mass, and spin can now be now delivered in less than half the time and with better results than one of the world's experts, with potential for further improvements. [44].
- Applied machine learning to determine the most effective asteroid deflection technology – a "Deflector Selector" based on over 1.5 million impactor scenarios. This new approach provided a tool to determine efficacy of various technologies to deflect or disrupt an Earth-affecting asteroid, delivering a successful deflection rate of 98% based on recommended approaches. [45][46].
- Submission of three authored papers and two posters accepted at the American Astronomical Society Division for Planetary Sciences meeting in October 2016, the International Academy of Astronautics 2017 Planetary Defense Conference, and numerous AI conferences including Nvidia's GPU Technology Conference 2017.

The progress made by the three teams was deemed of high potential utility to the planetary defense community, while also pushing the boundaries of machine learning methodologies and application. Jen-Hsun Huang, CEO of Nvidia said, "We now have a powerful new tool to tackle some of the world's greatest challenges. Finding potentially dangerous asteroids is surely one of those challenges - and this one is potentially existential."

Based on positive prospects for the FDL format, public-private partnership opportunities, and AI as a keystone in scientific progress, NASA supported an expanded initiative through a second cohort in 2017 with five interdisciplinary teams and 24 post-graduate professionals. The 2017 FDL also featured increased scope and broadened partner engagement with IBM, Lockheed Martin Corporation, Kx

Systems, Intel, and SpaceResources Luxembourg joining Nvidia and SETI with expanded financial and in-kind support. The second FDL built on the investigation of promising results from the first, specifically the radar shape modeling activity. The overall program continued to focus on planetary defense, with specific attention on long-period comets and shape modeling of the asteroid population using AI radar-inversion methods. New topics included lunar water and volatiles, solar storm prediction and solar terrestrial interactions [47]. At the time of this paper, the teams were in the process of completing technical papers and preparing for conference presentations.

A significant result from FDL's second cycle was the level of support from IT industry leaders - both in-kind (specifically GPU computing resources and AI expertise) and capital provisions. As an early example of new public-private partnerships, Intel, Nvidia, IBM, Lockheed Martin, Kx Systems, AutoDesk Inc., and Google, contributed staff, vast computing resources, and funds to FDL - enabling its researchers to iterate and experiment quickly with the large data sets that accompany NASA challenges. This ability to replicate quickly, test various approaches, gather lessons learned, and leverage expertise from industry is a key strength of FDL.

These companies recognized the value of an association with NASA, and in assisting the agency with seeking non-traditional solutions to essential societal problems, such as space-weather prediction and asteroid characterization. Correspondingly, the project benefits from the wisdom and guidance of world leaders in AI.

FDL is an example of an ideal partnership, where each partner accomplishes more than would be possible independently, to resolve difficult problems for the benefit of humanity. Industry partners receive the opportunity to experiment with their technology in a real-world setting, contribute to challenging and meaningful problem solving, and work with world-class research talent. In some cases, these efforts have led to employment offers for FDL participants. NASA leveraged its investment to enable a program beyond its budgetary capacity, and gained access to cutting-edge machine learning technology that in many cases was unavailable to the public. The private sector partners fully funded the program expenses of all the foreign participants, as NASA is precluded from spending taxpayer dollars on foreign nationals. This arrangement ensures that the best students in the world have the opportunity to participate in FDL.

At the time of the writing of this paper, a third year of FDL was in planning stages, and featured a broader expansion to include Earth observation, astrobiology, and orbital debris as potential research areas, as well as application of emerging AI tools such as reinforcement learning and quantum AI. Also at the time of he writing of this paper, the supporting scientific paper discussing the implications of the FDL 'Deflector Selector' project was accepted by the journal Acta Astronautica—becoming the first peer-reviewed paper to emerge from FDL related activities. FDL is a proven model of collaboration and problem solving beyond its origins in planetary defense, however the project retains a prime focus on efforts with great collective benefit to society - becoming the most impactful of all the AGC projects.

## 6. AGC Results and Lessons

The origins and results of the AGC from 2012-2017 provide lessons in strategy, policy, and project selection and execution. This section summarizes the project level impacts of the AGC program over five years, the strategy and policy tools that were essential in enabling NASA to use these approaches in the first place, and observations about the overall success to date of the grand challenge.

NASA initially aimed to establish early AGC successes supporting the premise that open innovation and public engagement methods combined with traditional investments and scientific community engagements are effective in delivering mission results and accelerating progress. Therefore, during the effort's early stages, OCT focused on areas most likely to benefit from open innovation solutions while the NEOO

Program generally focuses on traditional investments. However, as the AGC effort matured and public inputs were incorporated, NASA identified additional areas for projects with higher potential impact. A number of these activities provided measurable results with near-term impact in planetary defense, e.g. the FDL. Other activities were less impactful or could not be completed, though providing experiential lessons that can be considered for future program planning.

The AGC's efforts focused primarily on the improvement of detection and follow up observations to ensure the availability of adequate warning of immediate threats. Therefore, many of the projects supported through the AGC did not address the grand challenge's second component, to find all asteroid threats to human populations and know what to do about them. "Know what to do about them" is simple phrasing for the concept that these disasters can be prevented with adequate warning and development of space-based technologies. Technically, this element of the challenge is more difficult to support meaningfully with open innovation and public engagement. Despite this, one of the three FDL projects from year one utilized machine learning to develop a 'Deflector Selector' tool to analyze which deflection technique would be best used under different asteroid variables. Also, the ECAST public engagement and the FDL directly addressed this aspect. In the case of ECAST, the participants were asked to assess various technologies that would be required to deflect or disrupt an incoming asteroid. The ECAST project yielded particularly valuable data and insights.

This final section of the paper describes the important legal and policy authorities that enabled NASA to pursue this grand challenge approach, a set of lessons learned, and a high-level assessment of the effectiveness of the AGC.

*6.1. NASA'S Foundational Strategic and Policy Support of Grand Challenges*

Many of the projects supported through the AGC would have been impractical without the unique strategy and policy foundation NASA had in place to support this initiative. NASA laid the groundwork to support agency grand challenges by developing supportive policy and leveraging approaches like prize competitions and non-traditional partnerships as major vehicles for implementation. The AGC gave NASA's existing audiences a new way to connect actively with NASA on a new project, and opened channels to previously unengaged audiences. For agencies that do not have foundational strategy, authorizing legislation and policy in place supportive of these engagement approaches, some precursor work might need to occur before conducting a public participation based Grand Challenge.

Existing strategy and policy support structures at NASA that enabled the agency to pursue partnerships and innovative approaches included:

- Space Act Agreements (SAAs): The National Aeronautics and Space Act of 1958 provides NASA with the unique mandate to share its mission with various stakeholders, and the authority to enact a wide range of "other transactions," including SAAs. While NASA often utilizes partnerships to enhance its work, these strategic collaborations are considered fundamental to the AGC.

- Challenges, Prize Competitions, Crowdsourcing and Citizen Science: Since 2005, under the Prize authority section of the National Aeronautics and Space Act (42 U.S.C. § 2459f-1), the NASA Centennial Challenges Program has conducted numerous prize competitions, including: the Tether Challenge, the Beam Power Challenge, the Astronaut Glove Challenge, the Lunar Lander Challenge, and the Cubequest Challenge. NASA's flagship prize program has demonstrated the value these competitions can provide to agency technology development and demonstration efforts. Building on the success of the Centennial Challenges program, increased congressional support of these approaches, and other agency crowdsourcing efforts, in 2014 NASA "codified" challenges as a recommended agency tool by initiating NASA Policy Directive (NPD) 1090.1 Challenges, Prize Competitions, and Crowdsourcing Activities [48]. This policy also encourages managers to utilize innovative approaches to scientific research by engaging citizen solvers (also known as citizen science) and developing open hardware platforms to scale and democratize discovery. While NASA use of challenges, prizes, crowdsourcing, and citizen science has continually expanded in cross-cutting fashion, these innovation tools are fundamental program elements of the AGC.

- NASA Strategic Plan: The 2014 strategic plan supported the premise of addressing challenges through public engagement [49], also highlighted the importance of planetary defense [50], providing the strategic support for a new grand challenge focused on asteroids.

## 6.2 Lessons Learned and Areas for Future Study

There are numerous areas for future study given the lessons from NASA's first grand challenge. Section 6.2 explores, in detail four primary lessons learned for future designers of public engagement grand challenges:
1. Understanding How and When to Engage Non-Experts
2. Understanding How to Leverage Grand Challenges to Build Coalitions and Respond to Other Technology Trends
3. Understanding How to Structure NASA Grand Challenges
4. Understanding Scientific Discovery Grand Challenges Overall

## 6.2.1 Understanding How and When to Engage Non-Experts

A major conclusion of the AGC is the non-trivial nature of the science and technical capabilities or resources needed to contribute materially to a wide-reaching and complex challenge, like defending our planet from asteroids. Amateur astronomers (i.e., the public) face significant challenges when attempting to contribute effectively to tracking and/or characterizing NEAs. To understand the barriers, it is worthwhile to understand the opportunity: a global pipeline of asteroid data is constantly monitored to identify asteroids on an impact trajectory with Earth, and observatories contributing data are expected to comply with rigorous quality control requirements. While several amateur astronomers (with expensive and 'professional-grade' home observatories) regularly contribute data, any other non-professional participant is also expected to meet these requirements. This requires adoption of a new and advanced skill set, in addition to the cost of a personal observatory capable of taking the measurements. The AGC underestimated this obstacle.

In addition, early champions of the AGC often envisioned a simple, quick pathway to participation for non-experts. However, the AGC was hampered by a lack of appreciation that early and ongoing engagement with the science community to identify realistic opportunities for open innovation would be necessary for the ultimate results to be embraced by that community.

As observed , George Washington University (GWU) researchers Szajnfarber and Vrolijk who studied the AGC during its active phase, "while NASA intended the [AGC] to complement the ongoing planetary sciences work, initial interactions with the [science] community were met with aggressive resistance— mainly due to their skepticism of the solvers and their professional identity destruction…the image that the AGC painted of an army of citizen scientists making backyard discoveries with hobby telescopes was, in the mind of the planetary science community, unrealistic. As such, external, nontraditional contributions [would] need to be focused on carefully selected sub-problems [51]." Another issue was the scientific community's unvoiced expectation that crowd-sourced algorithms and tools would be peer-reviewed before adoption by the asteroid observer expert community. The Asteroid Data Hunter Challenge and FDL projects both yielded what were considered successful tools but  are unused by NASA-funded observatories and will remain so until peer-reviewed work has been published. The FDL is actively working to fill this gap by encouraging the summer challenge teams to continue their collaboration until a technical paper can be published in a relevant journal. A prior engagement with the expert stakeholder community on the value of crowd sourcing and innovation platforms would also help ensure adoption by that community.

Appreciating the need for more intentional collaboration with the scientific community after repeatedly facing these barriers, NASA and GWU conducted a workshop in July 2015 to identify the opportunities for open innovation with the planetary science community. This workshop yielded two areas that the science community believed could potentially be addressed through open innovation, including inferring the shape of known bodies (a characterization task) and harnessing the knowledge of the geology and seismology communities for new measurement techniques and instrumentation for generalizing composition characteristics [52]. As noted by Szajnfarber and Vrolijk, "the AGC team had spent nearly two years seeking to gain traction on a set of problems where the planetary science community would value open input. This 2-hour workshop was successful in that it made meaningful progress towards that goal [53]." A major lesson learned for future scientific discovery grand challenges is to conduct a facilitated problem decomposition workshop with the relevant science community much earlier in the process.

Understanding the subproblems that the AGC and science community agreed would be high value, several partner organizations were engaged to build pathways for the general public to develop the skills needed to participate.  Figure 2 highlights a skill curve for contributions to the AGC, showing where different dimensions of the problem were assigned to a spectrum of contributors. They range from empowering existing professionals, and engaging previously uninvolved but highly skilled experts in academia (e.g. Frontier Development Lab), to building awareness and pathways for pro-amateurs (amateurs with professional levels of skill) and leveraging the largely untapped global community of passionate citizen scientists and makers. Tools such as the Ultrascope and Verizon's asteroid characterization game were developed for this last group, who have the capacity to understand and lend their time and talent but lack the means.

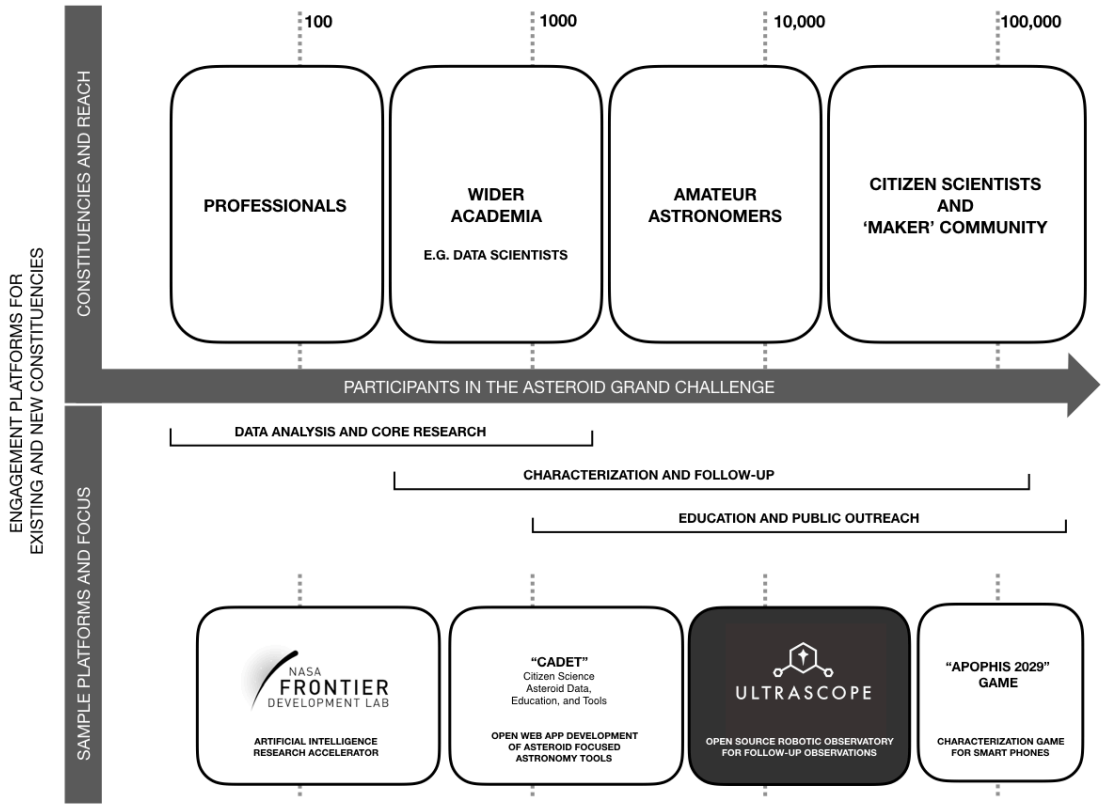

**Figure 2: Skill Curve for Contributions to the AGC**

The AGC was able, ultimately, to identify several subproblems to which non-experts and the general public could meaningfully contribute. By engaging broader groups, AGC successfully amplified its reach as an education and public outreach program, however in order to increase the likelihood of adoption of the solutions within the scientific community, more intentional engagement with the planetary science community would have been beneficial. To maximize impact, future AGC activities should organize projects based on task difficulty and the target audience. The FDL is an example of a later-phase AGC activity that considered these lessons in its design, with positive results.

*6.2.2 Understanding How to Leverage Grand Challenges to Build Coalitions and Respond to Other Technology Trends*

Grand Challenges are also a means to build broad multi-stakeholder coalitions and provide a mechanism for organizations to leverage their skill and ability in solving problems larger than they can successfully address on their own. In this way, they remain a viable framework for tackling many of this century's toughest problems, which are characterized by systemic intransigence and market failure. This can be particularly useful for government agencies as funding can be limited and particular problems can benefit from leveraging the expertise and resources from the private sector. For example, without large upfront financial commitments, the United State Agency for International Development (USAID) has energized diverse partnerships and on a wide range of topics through grand challenges [54]. However, it is helpful to engage a foundational community already interested or focused on the problem, where the grand challenge can focus the attention of a motivated population, providing inspiration, momentum, and a sense of belonging.

NASA's global reputation means that the agency is particularly well suited to multi-stakeholder leadership - which are often aptly referred to as "Moonshots". In the future, potential NASA grand challenges should seek problems with significant private sector involvement or interest. In the case of the AGC, the planetary defense community is a small subset within existing planetary research funding, innovating independently for years due to continual underfunding. Prior to the AGC, little new commercial interest or engagement occurred on this topic, with planetary defense seemingly a problem of commitment and not expertise.

In addition to multi-stakeholder engagement, Grand Challenges offer a method for exploring the potential of exciting new waves of technology that could be considered part of a formula for success. By pairing an inspiring and impactful vision with emerging technologies, like those in 2014 and 2015 which included low-cost additive manufacturing, cloud applications, and artificial intelligence--grand challenges are bolstered by the movements and corporate resources that inherently follow. The Ultrascope and FDL projects were examples of projects specifically designed to leverage emerging technologies to address the challenge. Moreover, new generations of graduates often establish themselves by embracing emerging technologies (such as the maker movement and the current proliferation of data scientists) providing the motive and means to apply their ingenuity to intractable global problems.

### 6.2.3 Understanding How to Structure NASA Grand Challenges

When the AGC was established by NASA in 2013, its associated open innovation activities were coordinated from OCT with one full time employee and a $900k annual budget, a small fraction of the agency's annual $17B budget. These activities were executed in close coordination with the SMD NEOO Program. STMD and HEOMD maintained interests in the AGC, although their budgetary responsibilities primarily were linked to ARM. In terms of AGC projects, HEOMD played a primary role through the NASA Tournament Lab with the Asteroid Data Hunter Challenge, and led the development of the SAA with Planetary Resources, while STMD led the CubeQuest Challenge. Because so many organizations could contribute to the open innovation components of the AGC, OCT—the senior technology advisory organization to the NASA Administrator—was selected as the coordinating organization. While this separation allowed for more aggressive open innovation experimentation than would have been possible if the AGC had been initially assigned to SMD, this separation drew a clear line between the roles and responsibilities for AGC public engagement activities and the scientific community's AGC involvement through the NEOO Program. Also, due to the small budget controlled by OCT, the Program Executive for the AGC led by influence rather than authority, to design and implement new activities that would complement SMD, HEOMD and STMD efforts and leverage existing investments.

Choosing a first grand challenge that required coordination across three of NASA's four Mission Directorates added significant complexity. Integration across Mission Directorates within NASA is typically difficult, and more so when implementing an entirely new way of structuring projects and programs. For future grand challenges, the authors recommend that NASA consider assigning responsibility for its open innovation and public engagement elements to the organization responsible ultimately for the overall science and technology goals: in the AGC's case, this organization is SMD. If expertise with open innovation and public engagement methods within that organization is lacking, NASA could assign appropriate staff to those roles within the organization. This would increase the probability that traditional and open innovation methods would be more seamlessly integrated. Otherwise, efforts that advance public participation and achievement of specific agency goals can conflict.

Embracing new approaches like open innovation requires shifts in mindset and the way that scientists and technologists collaborate to achieve different parts of their mission. Given the pace of work, managers often are unable to begin considering utilizing open innovations methods. Without considering these engagements and methods holistically from a program level and assigning responsibility to the mission organization, open innovation activities could be easily dismissed by a technical organization as "education and outreach". The AGC was regarded as such many times within NASA and the planetary science community. However, the AGC has shown that some promising ideas and contributions were made that would have been unlikely from NASA's traditional performers.

When designing a grand challenge, organizations should carefully determine the most appropriate location within the organization for the leadership and management of that challenge. For NASA, co-locating the AGC within SMD along with appropriate open innovation and public engagement staff, could have enabled closer ties with the NEOO Program, earlier collaboration with the planetary sciences community, and a vested interest in the AGC's success.

*6.2.4 Understanding Scientific Discovery Grand Challenges*

A scientific challenge with societal implications, the AGC sought to accelerate progress by infusing new thinking, audiences, and approaches into the traditional manner of addressing a problem. NASA's experience with the AGC reveals a number of questions for future study related to successfully implementing grand challenges that seek to engage non-traditional stakeholders in the advancement of scientific discovery and understanding.

Not all grand challenges are alike. The numerous types of grand challenges include those focused on scientific discovery or technology adoption. Future efforts must differentiate between the types of grand challenges that are pursued, since this understanding and program design may influence the assessment of the technological resources and commitment required to achieve associated objectives. The AGC reveals this is a vital factor in designing the challenge goal itself. By examining AGC's choice of framing ("to find all asteroid threats to human populations and know what to do about them"), both the leading and ending portions of the statement state two distinct scientific purposes for the challenge. Although meant to be inspirational, this framing is simultaneously disempowering, as ultimately only a handful of governments have the resources to observe dim objects that threaten cities and human life, or respond to and identified threat. Moreover, little incentive exists for non-government organizations or non-experts to invest their energies as a business or professional venture, since a future market for asteroid detection and redirection has yet to be established. This essential problem encumbered the AGC with a core dilemma: how to engage new constituencies with an extraordinarily challenging problem that seemed only within the reach of governments and scientists, compounded with a lack of definitive rewards for taking the risk.

Furthermore, other grand challenges have better understood the role of government to kick-start an initiative to investigate a problem with potential for democratization and a future commercial opportunity, creating a pathway for other actors to follow. For example, the global effort to sequence the entire human genome involved a $3.8B investment by the U.S. Federal Government over 13 years [7]. Another example of a scientific grand challenge, the ongoing BRAIN initiative launched in 2013, seeks to deepen understanding of the inner workings of the human mind and to improve how disorders of the brain are treated, prevented, and cured. As of December 2017, the NIH has made at least four rounds of awards for BRAIN initiative-related research, and there were 30 open and 90+ closed opportunities for funding from the U.S. Federal government (NIH, National Science Foundation, Intelligence Advanced Research Projects Activity, DARPA) and non-government partners (Simons Foundation, Kavli Institute for Theoretical Physics). To date, hundreds of millions of research dollars have been awarded to researchers to advance the goals of the BRAIN initiative [55].

Instead of selecting a future challenge that is focused on scientific discovery, NASA could consider efforts that accelerate the pace of technology development or deployment. This will enable an assessment of how different types of Grand Challenges produce different types of results in the aerospace sector.

*6.3 Reflections on the Success of the AGC*

Despite the challenges of engaging non-experts in the AGC since its announcement in 2013, the budget for the overall NEOO Program and PDCO were increased substantially and sustained [56] across two administrations. From 2013-2017 the modest budget for the AGC was not increased and the AGC was discontinued in October 2017. During the AGC's implementation by OCT, it was considered a complementary experiment that could contribute to the success of NASA's NEOO Program goals. However, when the AGC was transferred to SMD it was not included in the NEOO Program's ongoing budget development. While the PDCO continues to support some existing AGC projects such as FDL, dedicated AGC funding is no longer available as of Fiscal Year 2018.

Given that the AGC has concluded, its success criteria can be described in three ways: at the grand challenge characteristic level; at the individual project level, with respect to the initial "find, track, characterize and mitigate" grand challenge goals; or with respect to the iterated grand challenge goals of "awareness, follow up, and characterize".

The EC characteristics of a NASA Grand Challenge were defined in section 2 of this paper. By those standards, the AGC could be seen as largely successful in three of four of the characteristics:
1. The AGC clearly *related to and continues to relate to NASA's core mission and strategy*, with annual agency budgets continuing to support the efforts to find and protect against asteroid threats.
2. The AGC also made progress towards *unifying and inspiring public and private constituencies around a clear "call to action"*. NASA effectively deconstructed the overall problem statement to a few subproblems where non-experts could be directed to make meaningful contributions to a call to action. Furthermore, NASA directed attention to a global challenge and then channeled the new awareness through non-traditional opportunities for public participation, resulting in hundreds of individuals engaging in the AGC depending on the project.
3. The AGC also served as a *beacon for high-impact, multi-disciplinary collaborations among the government, industry companies, universities, non-profits, philanthropists, and individuals including scientists and engineers*. A number of new formal and informal partnerships emerged as a result of the AGC that may not have been pursued otherwise: for example, the SAA with SpaceGAMBIT and Verizon as well as the ongoing engagement of Intel and Nvidia with the FDL.
4. The characteristic that has been less clearly achieved to date is how the AGC *spurred significant developments in a wide range of domains vital to the nation and the world.* The authors do believe that the solutions and approaches furthered through projects such as CubeSat technologies developed in the CubeQuest challenge, the Ultrascope, and FDL solutions hold significant potential. However, insufficient time has passed in order to assess the long-term impact of these solutions.

In the end, the AGC found opportunities for non-traditional groups to participate in a meaningful way to "find all asteroid threats to human populations and know what to do about them". In a declining budget environment, experimental activities such as the AGC require monitoring and independent assessment to justify continued funding past the initial phase. Due to its nature as a challenge, the AGC was not intended to continue indefinitely. However, SMD did consider was whether the results of the specific public participation-oriented projects merited continued funding.

For example, SMD determined that some projects did not merit additional funding as the results were inconclusive or newly engaged participants were unable to continue or expand promising activities despite attempts to encourage them (e.g. the badging project and Verizon collaboration). For approaches that did show promise but that were unable to be continued due to changing partner commitments, (e.g. the badging project and Verizon collaboration), the authors recommend that NASA consider re-invigorating those projects if funding becomes available. Also, a number of projects developed through the AGC successfully engage non-traditional communities, leverage technology trends outside asteroid science, and continue to be supported by SMD. For example, as illustrated in the case study section of this paper, FDL has the distinct potential to yield successes.

Finally, the external interest and enthusiasm that NASA inspired by announcing and coordinating a grand challenge is encouraging. An informal Facebook group with over 2000 followers was created by participants in multiple AGC activities, and remains active despite no active involvement from NASA [57]. NASA effectively directed attention to a global challenge and then channeled the new awareness through non-traditional opportunities for public participation, resulting in over a million individuals and organizations interacting with the AGC through the projects described in this paper. Communities with interest in contributing to space exploration perceived an accessible avenue for participation, and the response from these non-traditional stakeholders has been positive. By engaging the public through the AGC, NASA created science, technology, engineering, and mathematics engagement and outreach opportunities; and stimulated the awareness needed to defend our planet from the threat of asteroids – all at minimal cost to the taxpayer.

## 7. Acknowledgements

The authors wish to thank Lori Garver, Mason Peck, Rebecca Spyke-Keiser, Cristin Dorgelo, and Tom Kalil who were critical to AGC concept development, early champions of the approach, and valuable partners in developing the foundational AGC strategy. We also thank the thought leaders who have helped evolve knowledge and understanding related to the AGC since its announcement on June 18, 2013: Jose Luis Galache, Jeff Hamaoui, and Blair Walter Tom. Recognizing the value in learning from the methodology of the AGC, we engaged George Washington University with a grant to Zoe Szajnfarber and Ademir Vrolijk. They analyzed the AGC process to facilitate an unbiased scientific analysis and contribution to literature on the expected value of the approach. The internal NASA progress of the AGC would not have been possible without the help and support of colleagues throughout the agency, particularly: Lindley Johnson in SMD, Michele Gates and Jason Crusan in HEOMD, and James Reuther in STMD. Several projects described i this article also would not have been possible without the contributions of NASA colleagues Amy Kaminski, Zachary Pirtle, Dan Rasky, and Chris Moore.